\documentclass[letter,twocolumn]{jpsj3}
\usepackage{graphicx,bm,times,amsmath}
\allowdisplaybreaks
\newcommand{\I} {\mathrm{I}}
\newcommand{\II}{\mathrm{I\hspace{-.1em}I}}
\def\runauthor{M. Tsuchiizu \textit{et al.}}

\title{
Multi-Orbital Molecular Compound (TTM-TTP)I$_3$: \\
Effective Model and Fragment Decomposition 
}

\author{
Masahisa \textsc{Tsuchiizu}%
$^{1}$\thanks{E-mail: tsuchiiz@s.phys.nagoya-u.ac.jp}, 
Yukiko \textsc{Omori}$^{1}$, 
Yoshikazu \textsc{Suzumura}$^{1}$,
Marie-Laure \textsc{Bonnet}%
$^{1,2}$\thanks{%
Present address:
Institute of Physical Chemistry, University of Zurich,
Winterthurerstrasse 190, 8057 Zurich, Switzerland.}, 
\\
Vincent \textsc{Robert}$^{2,3}$, 
Shoji \textsc{Ishibashi}$^{4}$, 
and
Hitoshi \textsc{Seo}$^{5,6}$
}

\inst{$^{1}$%
Department of Physics, Nagoya University, Nagoya 464-8602, Japan
\\
$^{2}$%
Universit\'e de Lyon, Laboratoire de Chimie,
Ecole Normale Sup\'erieure de Lyon, 
CNRS, F-69364 Lyon, France
\\
$^{3}$%
Laboratoire de Chimie Quantique,
UMR 7177 CNRS/Universit\'e de Strasbourg,
F-67000 Strasbourg, France
\\
$^{4}$%
Nanosystem Research Institute (NRI) "RICS",
AIST,
Ibaraki 305-8568, Japan
\\
$^{5}$%
Condensed Matter Theory Laboratory, RIKEN, Wako, Saitama 351-0198, Japan
\\
$^{6}$%
JST, CREST, Wako, Saitama 351-0198, Japan
}

\recdate{December 27, 2010: J.\ Phys.\ Soc.\ Jpn. \textbf{80} (2011) 013703}

\abst{ 
The electronic structure of the molecular
compound (TTM-TTP)I$_3$, which exhibits a peculiar intra-molecular
charge ordering, has been studied using multi-configuration {\it ab initio}
calculations.  
First we derive 
an effective Hubbard-type model 
based on the molecular orbitals (MOs) of TTM-TTP; 
we set up a two-orbital Hamiltonian 
for the two MOs near the Fermi energy 
and determine its full parameters:
the transfer integrals, the Coulomb and exchange interactions. 
The tight-binding band structure 
obtained from these transfer integrals 
is consistent with the result of the
direct band calculation 
based on density functional theory.  
Then, by decomposing the frontier MOs into two parts, i.e., fragments, 
we find that the stacked TTM-TTP molecules
can be described by a two-leg ladder model, while the inter-fragment Coulomb 
energies are scaled 
to the inverse of their distances. 
This result indicates that the fragment picture that we proposed earlier
[M.-L. Bonnet {\it et al.}: J.\ Chem.\ Phys.\ \textbf{132} (2010) 214705]  
successfully describes the low-energy properties of this compound. 
}

\kword{
multi-orbital molecular compound, 
intra-molecular charge ordering,
\textit{ab initio} calculation,
band calculation, 
fragment decomposition
}
  
\begin{document}
\maketitle

A rich variety of physical properties have been 
explored in molecular solids 
since the realization of the first molecular metal,
TTF-TCNQ, and in the last decade,
charge-ordering (CO) phenomena and related issues have attracted
much attention. \cite{Mori2004,Seo2004} 
Despite the complexity of the constituent molecules, 
a simple picture works well for many compounds;
the low-energy electronic properties 
are determined by a single frontier molecular orbital (MO). Therefore 
each molecule can be regarded as a single site, 
and then the tight-binding and Hubbard-type models 
based on this 
picture successfully describe
their physical properties. 
\cite{Seo2004,Seo2006}
However, it has recently been recognized that 
 such a simple single-MO
approximation is not sufficient for
 describing the electronic structures of 
 single-component molecular metals $M$(tmdt)$_2$ [$M=$ Ni, Au, etc.]; 
\cite{Tanaka_SCMM2001,Rovira,Ishibashi-Seo-SCMM} 
 there are cases where
 multi-orbital effects reflecting the respective MOs
 induce unexpected novel physical properties.

In the present paper, we focus on a
quasi-one-dimensional
molecular crystal (TTM-TTP)I$_3$.
\cite{Mori1994,Mori1997}
Since the formal charge of the
TTM-TTP molecule is $+1$ owing to the monovalent counterion I$_3^-$, this
compound was considered to be a half-filled system and the possibility of 
a genuine one-dimensional Mott insulator with paramagnetic spin properties was
discussed. \cite{Mori1997,Yasuzuka2006,Tsuchiizu2007}
However, it has also been pointed out that
this system exhibits a transition toward a nonmagnetic
state at around 80~K.  \cite{Maesato1999,Fujimura1999,Onuki2001} 
On the basis of detailed experimental analyses of this low-temperature
 phase by Raman scattering \cite{Yakushi2003} and 
 X-ray measurements, \cite{Nogami2003}
a new type of CO state has been proposed.
This state is called the 
``intra-molecular CO'' (ICO)  state, 
where the inversion center on the middle point of the TTM-TTP
molecule is lost and the charge is disproportionated 
\textit{within} each molecule.
This state cannot be described by the simple single-MO picture by its nature.

\begin{figure}[t]
\begin{center}
\includegraphics[width=7cm]{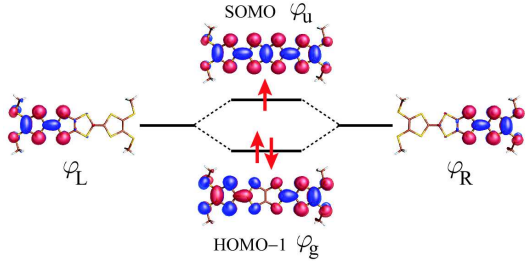}
\end{center}
\caption{
(Color online)
SOMO and HOMO-1 of the ionic [TTM-TTP]$^+$ molecule
obtained by the restricted 
open-shell Hartree-Fock procedure,\cite{Marie2009} and 
a schematic energy diagram based on the left and right fragment MOs.
}
\label{fig:fig1}
\end{figure}

A key to understanding the electronic state of this compound, 
 as shown in our previous theoretical work,\cite{Marie2009} 
 is that 
 the singly-occupied-molecular-orbital (SOMO) and 
 the second-highest-occupied-molecular-orbital (HOMO-1) 
 of the ionic [TTM-TTP]$^+$ molecule
 are
close in energy. 
The chemical reason of this quasi-degeneracy 
 can be understood in terms of a three-fragment model. \cite{Marie2009}
The SOMO and HOMO-1 are basically described by 
 two of the fragments, as in the following. 
Since the TTM-TTP molecule itself 
 has an inversion center,
 the resulting MOs can be classified into
  gerade (g) and ungerade (u) MOs.
The SOMO (u MO) and HOMO-1 (g MO) are shown in 
Fig.\ \ref{fig:fig1}, where the SOMO 
  exhibits a bonding character between the left and right fragment MOs 
while the HOMO-1 has an 
antibonding character, i.e.,
\begin{equation}
\varphi_{\mathrm u} 
\approx
\frac{1}{\sqrt{2}} 
\left(
  \varphi_{\mathrm L} + \varphi_{\mathrm R}
\right)
,\quad
\varphi_{\mathrm g} 
\approx
\frac{1}{\sqrt{2}} 
\left(
 -  \varphi_{\mathrm L} + \varphi_{\mathrm R}
\right),
\label{eq:fragment}
\end{equation}
where $\varphi_\mathrm{L}$ and $\varphi_\mathrm{R}$
are the left (L) and right (R) fragment MOs.   \cite{Marie2009}
The purpose of the present paper is to construct 
 a minimal model 
 that can describe the ICO state.
 The contribution of the center (C) fragment MO 
  can be  neglected for this purpose, since 
  the C fragment is deep in energy. 
  \cite{Marie2009}
We  construct a two-orbital Hubbard-type
 model based on the SOMO and HOMO-1.
 The magnitudes of the intra-molecular and inter-molecular interactions 
 are estimated from 
 \textit{ab initio} calculations with a complete-active-space
 configuration-interaction (CAS-CI) method. \cite{MOLCAS} 
By transforming the model into the fragment picture, we clarify that
   the stacked TTM-TTP molecules can be described by a 
 two-leg ladder model.

The \textit{ab initio}
  calculations were performed using the Molcas7 package
\cite{MOLCAS}
  with the basis set:
 S(7s6p1d)/[4s3p1d], C(5s5p1d)/[3s2p1d], and  H(3s)/[1s].
The atomic parameters are taken from the results of 
the X-ray structure analysis at room temperature \cite{Mori1994}.  
As mentioned in ref.\ \citen{Marie2009}, the SOMO and HOMO-1  
 levels are well separated from the other MOs, 
 and therefore, these two MOs are 
used to generate CAS[3,2] containing three electrons
in two MOs.
In order to evaluate the magnitudes of the inter-molecular interactions,
we select two neighboring molecules and, in this case, put four MOs into the
active space.
Throughout the CI calculation, 
the MOs are fixed to the ones obtained from the 
  restricted open-shell Hartree-Fock calculation.

First we derive the effective  model Hamiltonian 
  describing the isolated 
[TTM-TTP]$^+$ ion. 
In the CAS[3,2], there are two independent configurations:
  (i) one is the ungerade state $\mathrm g^2 \mathrm u^1$ 
in which two electrons are on 
  g  and one electron is on u, 
  and (ii) the other is the gerade state $\mathrm g^1 \mathrm u^2$  
with one electron on  
  g  and two electrons on u.
In order to evaluate the Coulomb interactions, we also consider the
  configurations
 arising from the occupations of g and u MOs by 0, 1, 2 and 4 electrons.
By the CAS-CI \textit{ab initio} calculations,
  we can obtain the full energy spectrum and the 
information of the wave functions.
By collecting these data, we can construct the two-orbital 
Hubbard-type Hamiltonian
  that reproduces the energy spectrum.
Namely, 
we consider the following one-molecule Hamiltonian:
\begin{align}
 \hspace*{-.3cm}
H_{1\mbox{-}\mathrm{mol}}
=
& \, E_0
  + \varepsilon_{\mathrm{g}}^0 n_{\mathrm{g}}
   + \varepsilon_{\mathrm{u}}^0 n_{\mathrm{u}}
\nonumber \\
  &   
 + U_\mathrm{g}  n_{\mathrm{g},\uparrow} n_{\mathrm{g},\downarrow}
 + U_\mathrm{u}  n_{\mathrm{u},\uparrow} n_{\mathrm{u},\downarrow} 
 + U' n_{\mathrm{g}} n_{\mathrm{u}}
\nonumber\\
& -    
J_\mathrm{H}
\left[
  \bm{S}_{\mathrm{g}} \cdot \bm{S}_{\mathrm{u}}
  - \frac{1}{2} 
   \big(
     c_{\mathrm{g},\uparrow}^\dagger c_{\mathrm{g},\downarrow}^\dagger
     c_{\mathrm{u},\downarrow} c_{\mathrm{u},\uparrow} + \mathrm{h.c.}
  \big)  
\right]	,
\label{eq:H_1mol}
\end{align}
where $c_{\mathrm{g},\sigma}$ ($c_{\mathrm{u},\sigma}$) is the 
electron annihilation operator for g (u) MO 
  of the TTM-TTP molecule.
The quantities $\varepsilon_{\mathrm g}^0$ and 
$\varepsilon_{\mathrm u}^0$ represent the energy levels of the
   g and u MOs.
$U_{\mathrm g}$ ($U_{\mathrm u}$) represents the Coulomb repulsion 
  between two electrons on the g (u) MO, and 
 $U'$ is the inter-MO Coulomb repulsion.
 $J_{\rm H}$
 is the Hund coupling including the pair-hopping term,
and $E_0$ is the energy constant.
In contrast with the case of atomic orbitals under the centrosymmetric 
  potential, there is no constraint relation among these couplings.
The density operators are 
  normal ordered, i.e., 
  $n_{\nu,\sigma}=(c_{\nu,\sigma}^\dagger c_{\nu,\sigma}^{}-3/4)$
 with $\nu=\mathrm{g},\mathrm{u}$.  
Here, we note that the charges on each MO are set as $3/4$ 
  in order to treat the g and u MOs on the same footing. 
 \cite{Marie2009}
The evaluated parameters are summarized in 
Table \ref{table:parameter1}. 
The energy level difference between u and g MOs is 
relatively small, $\approx  0.42$ eV, 
which reproduces the results reported in ref.\ \citen{Marie2009}. 
The magnitudes of  $U_\mathrm{g}$ and $U_\mathrm{u}$ are small
compared with those of smaller molecules such as the TTF molecule in TTF-TCNQ,
for which we obtained $U_{\mathrm{TTF}}\approx 6.2$ eV.
In addition, $U_\mathrm{u}$ is slightly larger than $U_\mathrm{g}$, a
  reflection
of the left$-$right bonding character of the u MO  compared with 
the antibonding character of the g MO. 
The Coulomb repulsion for the TTF molecule
  was estimated from the density-functional-theory calculations 
to be $\approx 4.7$ eV, 
\cite{Cano-Cort2007}
 which is comparable to our result.
As a reference, 
we note that
the magnitude of the bare  Coulomb repulsion 
  of the BEDT-TTF dimer in the $\kappa$-(BEDT-TTF)$_2X$ system is 
  about $3$ - $4$ eV. \cite{Nakamura2009,Scriven2009} 
We also note that 
   relatively large Hund coupling 
  is obtained ($J_\mathrm{H}/U_\mathrm{u}\simeq 0.8$),
  in contrast to those
  for transition-metal atoms.

\begin{table}[t]
\caption{
Estimated parameters for the isolated TTM-TTP molecule.
}
\label{table:parameter1}
\begin{tabular}{ccc}
\hline\hline
Energy level  &
Intra-orbital interaction &
Inter-orbital interaction
\\
\hline
$\varepsilon^0_{\mathrm g}=-8.63$ eV &
$U_{\mathrm{g}}=3.70$ eV &
$U'=2.82$ eV
\\
$\varepsilon^0_{\mathrm u}=-8.21$ eV &
$U_{\mathrm{u}}=3.90$ eV &
$J_{\mathrm{H}}=3.19$ eV
\\
\hline\hline
\end{tabular} 
\end{table}

\begin{figure}[t]
\begin{center}
\includegraphics[width=8cm]{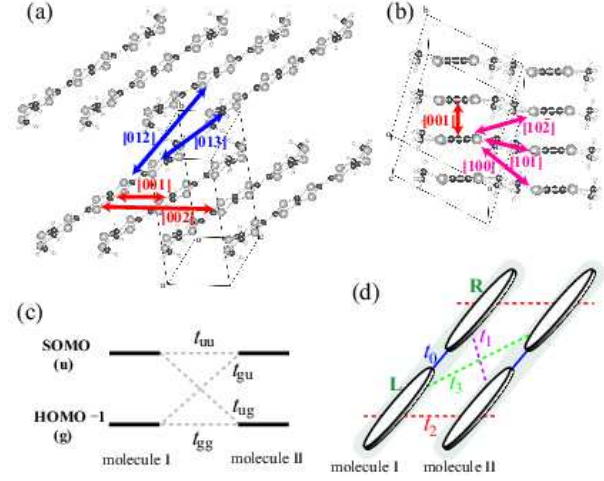}
\end{center}
\caption{
(Color online)
Crystal structure of the (TTM-TTP)I$_3$ projected 
onto the $b$$-$$c$ plane (a) and
  onto the $a$$-$$c$  plane (b).
Definitions of transfer integrals in the MO 
 basis (c) and in the
  fragment basis (d). 
}
\label{fig:fig2}
\end{figure}

Next we evaluate the inter-molecular interactions.
We focus on two neighboring TTM-TTP molecules 
   whose atomic coordinates are read from the 
  crystal structure at room temperature, and perform
   CAS-CI calculations
using the MOs for the isolated [TTM-TTP]$^+$ molecule.
In order to determine this set of MOs, 
the two molecules were artificially displaced 
with respect to one another so that the intermolecular overlap
was negligible.
Hereafter, we refer to the two target molecules as 
``molecule $\I$'' and ``molecule $\II$''.
The molecule pairs that we have focused on are shown in 
  Figs.\ \ref{fig:fig2}(a) and \ref{fig:fig2}(b).
The Hamiltonian for one-body terms that appear in the 
  two-molecule system is given by
\begin{eqnarray}
H_{2\mbox{-}\mathrm{mol}}'
\!\!\!\!\!\! &=& \!\!\!\!\!\!
- t_{\mathrm{gg}}
  c_{\mathrm{g},\I,\sigma}^\dagger c_{\mathrm{g},\II,\sigma}^{} 
- t_{\mathrm{uu}}
  c_{\mathrm{u},\I,\sigma}^\dagger c_{\mathrm{u},\II,\sigma}^{} 
\nonumber \\
&&  \!\!\!\!\!\! \hspace*{-.5cm} {}
-  t_{\mathrm{gu}} 
 c_{\mathrm{g},\I,\sigma}^\dagger c_{\mathrm{u},\II,\sigma}^{} 
-  t_{\mathrm{ug}} 
 c_{\mathrm{u},\I,\sigma}^\dagger c_{\mathrm{g},\II,\sigma}^{}
+ \mathrm{h.c.}
\nonumber \\
&&  \!\!\!\!\!\! \hspace*{-.5cm} {}
+  \Delta \varepsilon_{\mathrm{g}} 
   (n_{\mathrm{g},\I} + n_{\mathrm{g},\II})
+ \Delta \varepsilon_{\mathrm{u}} 
   (n_{\mathrm{u},\I} + n_{\mathrm{u},\II})
+ H'_\Delta,
\quad
\label{eq:Hkin-2mol}
\end{eqnarray}
where
the parameters $t_{\mathrm{gg}}$ and $t_{\mathrm{uu}}$ are the inter-molecular
   transfer integrals for the same g and u MOs, respectively,
 while 
  $t_{\mathrm{gu}}$ and $t_{\mathrm{ug}}$ are for different 
 MOs [see Fig.\ \ref{fig:fig2}(c)].
The summation over repeated spin indices is implied.
In terms of the
\textit{ab initio} Hamiltonian $H$, 
these transfer integrals are expressed as
$t_{\mathrm{gg}} \equiv
- \langle \varphi_{\mathrm g,\I} | H | \varphi_{\mathrm g,\II} \rangle$,
$t_{\mathrm{uu}} \equiv 
- \langle \varphi_{\mathrm u,\I} | H | \varphi_{\mathrm u,\II} \rangle$, and 
$t_{\mathrm{gu}} \equiv
- \langle \varphi_{\mathrm g,\I} | H | \varphi_{\mathrm u,\II} \rangle
=
+ \langle \varphi_{\mathrm u,\I} | H | \varphi_{\mathrm g,\II} \rangle
= - t_{\mathrm{ug}}$,
where
$|\varphi_{\mathrm{g},\I}\rangle$ ($|\varphi_{\mathrm{g},\II}\rangle$) and 
$|\varphi_{\mathrm{u},\I}\rangle$ ($|\varphi_{\mathrm{u},\II}\rangle$)
are the states with a single electron on 
the g MO and u MO in  molecule $\I$ ($\II$), respectively.
The parameters $\Delta \varepsilon_\mathrm{g}$ and
  $\Delta \varepsilon_\mathrm{u}$ represent the energy-level shift 
  due to the potential energy from the other molecule.
The additional term is 
$H'_\mathrm{\Delta}= \Delta\varepsilon_{\mathrm{gu}} 
  ( c_{\mathrm{g},\I,\sigma}^\dagger c_{\mathrm{u},\I,\sigma}
  - c_{\mathrm{g},\II,\sigma}^\dagger c_{\mathrm{u},\II,\sigma}
  + \mathrm{h.c.} )$,
  which reflects the asymmetry between L/R fragments within the
  molecule and  
   disappears when one considers the periodic crystal system.
The possible inter-molecular Coulomb interaction terms are represented as
\begin{eqnarray}
H_{2\mbox{-}\mathrm{mol}}''
\!\!\!\!\!\! &=& \!\!\!\!\!\!
       V_{\mathrm{gg}} n_{\mathrm{g},\I} n_{\mathrm{g},\II}
 +     V_{\mathrm{uu}} n_{\mathrm{u},\I} n_{\mathrm{u},\II}
\nonumber \\ && {} \hspace*{-1.5cm}
 +     V_{\mathrm{gu}} \left( n_{\mathrm{g},\I} n_{\mathrm{u},\II}
                         + n_{\mathrm{u},\I} n_{\mathrm{g},\II} \right)
\nonumber \\ && {} \hspace*{-1.5cm}
+
\Bigl[
 I   \,
(
   c_{\mathrm{g},\I,\sigma}^\dagger   c_{\mathrm{u},\I,\sigma}^{}
   c_{\mathrm{u},\II,\sigma'}^\dagger c_{\mathrm{g},\II,\sigma'}^{}
 + c_{\mathrm{g},\I,\sigma}^\dagger   c_{\mathrm{u},\I,\sigma}^{}
   c_{\mathrm{g},\II,\sigma'}^\dagger c_{\mathrm{u},\II,\sigma'}^{}
)
\nonumber \\ && {} \hspace*{-1cm}     
+
X_{\mathrm{g}} 
(
 n_{\mathrm{g},\I}
 c_{\mathrm{g},\II,\sigma}^\dagger c_{\mathrm{u},\II,\sigma}^{}
- 
 c_{\mathrm{g},\I,\sigma}^\dagger c_{\mathrm{u},\I,\sigma}^{}
 n_{\mathrm{g},\II}
)
\nonumber \\ && {} \hspace*{-1cm}     
+
X_{\mathrm{u}} 
(
 n_{\mathrm{u},\I}
 c_{\mathrm{g},\II,\sigma}^\dagger c_{\mathrm{u},\II,\sigma}^{}
-
 c_{\mathrm{g},\I,\sigma}^\dagger c_{\mathrm{u},\I,\sigma}^{}
 n_{\mathrm{u},\II}
)
       + \mathrm{h.c.} 
\Bigr],
\label{eq:H''}
\end{eqnarray}
%
\begin{table}[b]
\caption{
Estimated parameters for the inter-molecular interactions 
in the MO picture.
All energies are in eV. 
}
\label{table:intermol}
\begin{tabular}{c|rrrrrrr}
\hline\hline
 & [001] & [002] & [012] & [013] & [100] & [10\=1] & [10\=2]
\\ \hline 
$t_{\mathrm{gg}}$
& -0.04
& 0.00  
& 0.03
& -0.02
& 0.00
& 0.01
& 0.00
\\
$t_{\mathrm{uu}}$ 
& -0.29  
& 0.00
& -0.02
& 0.01
& 0.00
& 0.00
& 0.00
\\
$t_{\mathrm{gu}}$ 
& -0.13 
& 0.00
& -0.02
& 0.01
& 0.00
& 0.00
& 0.00
\\ \hline
$V_{\mathrm{gg}}$ 
& 2.12  
& 1.30
& 1.17
& 0.89
& 1.20
& 1.13
& 0.90
\\
$V_{\mathrm{uu}}$ 
& 2.30 
& 1.30
& 1.09
& 0.83
& 1.24
& 1.12
& 0.87 
\\
$V_{\mathrm{gu}}$ 
& 2.21  
& 1.31
& 1.13
& 0.85
& 1.22
& 1.13
& 0.89
\\ 
$I$ 
& 0.26 
& -0.03
& -0.14
& -0.09
& 0.06
& -0.05
& -0.06
\\ 
$X_\mathrm{g}$ 
& -0.27 
& -0.26
& -0.31
& -0.22
& 0.13
& 0.22
& 0.18
\\
$X_\mathrm{u}$ 
& -0.45
& -0.29 
& -0.29
& -0.20
& 0.17
& 0.23 
& 0.18
\\
\hline\hline
\end{tabular} 
\end{table}%
%
where $V_{\mathrm{gg}}$ ($V_\mathrm{uu}$) and $V_{\mathrm{gu}}$ 
  denote the inter-molecular density-density interactions 
  on the same g (u) MOs  and between the g and u MOs, respectively.
The $I$ term represents the orbital exchange interaction 
 and the $X_\mathrm{g}$ and $X_\mathrm{u}$ terms are the density-hopping
 interactions.
As in the case of the isolated molecule,
  we can determine the parameters in eqs.\ 
(\ref{eq:Hkin-2mol}) and (\ref{eq:H''}) by using 
the \textit{ab initio} energies 
  for states with different symmetry, spin states, and 
  electron numbers.
The evaluated parameters are summarized in Table \ref{table:intermol}. 
In contrast to the conventional 
 extended H\"uckel approach, 
 where the overlap integrals are used in the band calculation, \cite{Mori1984_Huckel}
 we can directly obtain the 
 transfer integrals 
 along different directions $[ijk]$
 in the present approach.
We find that only the nearest-neighbor 
transfer integrals along the stacking [001] direction are
  large and the magnitude of $t_{\mathrm{uu}}^{[001]}$
 becomes comparable to the 
  value estimated by the extended H\"uckel approach 
  ($\approx 0.26$ eV)\cite{Mori1994}.
We also note that the energy-level shifts are relatively large, e.g., 
  for the [001] molecule pair, 
 we find $\Delta \varepsilon_\mathrm{g}^{[001]}=-1.81$ eV and 
  $\Delta \varepsilon_\mathrm{u}^{[001]} =-1.85$ eV.
The magnitudes of the inter-molecular density-density repulsions are
  relatively large 
 compared with the intra-molecular interaction
  ($V_\mathrm{gg}/U_\mathrm{g}\approx 0.57$ and 
  $V_\mathrm{uu}/U_\mathrm{u}\approx 0.59$),
  whereas a similar value is also obtained in the benchmark
  TTF chain in TTF-TCNQ as  $V/U\approx 0.52$.

\begin{figure}[t]
\begin{center}
\includegraphics[width=7cm]{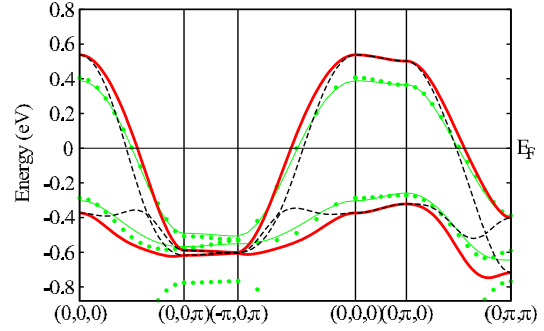}
\end{center}
\caption{
(Color online)
Band structure of (TTM-TTP)I$_3$ obtained 
from the tight-binding parameters evaluated by the CAS-CI 
\textit{ab initio} calculations
(thick  solid red lines).
The dashed lines represent the energy dispersion in the case of
  $t_{\mathrm{gu}}=0$.
The GGA and its fitted results are shown
  by green dots and thin solid green lines, respectively. 
}
\label{fig:fig3}
\end{figure}

On the basis of the results of the above analysis,
 we examine the band structure 
 of the (TTM-TTP)I$_3$ crystal
in the metallic state.
By neglecting correlation effects,
the  band structure 
 obtained from the crystal version of eq.\ (\ref{eq:Hkin-2mol})
is shown in Fig.\ \ref{fig:fig3}.
Here, we assign the energy levels for each MO as
  $\varepsilon_{\mathrm g} =
   \varepsilon_{\mathrm g}^0 +2\Delta \varepsilon_{\mathrm g}^{[001]} =
 -12.25$ eV and 
 $\varepsilon_{\mathrm u} =
  \varepsilon_{\mathrm u}^0 + 2\Delta \varepsilon_{\mathrm u}^{[001]} =
  -11.90$ eV.
 However, these absolute values should be refined by 
  taking into account other molecules as well as counterion I$_3^-$.
For comparison, 
electronic-structure calculations were carried out 
directly for the (TTM-TTP)I$_3$ crystal 
with the computational code QMAS (Quantum MAterials Simulator)\cite{qmas} 
based on the generalized gradient approximation (GGA).\cite{pbe} 
The resulting band structure is shown in Fig.\ \ref{fig:fig3}.
Despite the bandwidth being overestimated owing to the neglect of 
correlation effects, 
the overall band structure obtained from  eq.\ (\ref{eq:Hkin-2mol})
agrees with the GGA results, indicating the validity 
  of the present approach based on 
  the CAS-CI \textit{ab initio} calculations.
The hypothetical
band structure obtained by neglecting the SOMO$-$HOMO-1 mixing 
(i.e., we set $t_{\mathrm{gu}}=0$) is also shown by the dashed lines in Fig.\
  \ref{fig:fig3}.
We find that the SOMO band has large dispersion 
  along the [001] direction, whereas 
  the HOMO-1 band is very narrow and is located within
  the SOMO band.
Furthermore, these two bands are isolated 
   and thus we can conclude  that 
  this system can be regarded as a two-band system. 

Finally we transform the model into the fragment picture 
by using eq.\ (\ref{eq:fragment}), where
the stacked TTM-TTP molecules  can be described as
  an effective two-leg ladder system, as shown in Fig.\ \ref{fig:fig2}(d).
The  transfer integral between the L and R fragments 
 within the molecule is characterized by the energy difference 
 between the g and u MOs, i.e.,
$ t_0 = 
 (\varepsilon_{\mathrm{g}} - \varepsilon_{\mathrm{u}})/2
  \approx -0.17$ eV.
The Coulomb interaction within the fragment, i.e., 
  the ``on-site'' repulsion, is $U=5.30$ eV,
  which becomes comparable to that for the TTF molecule, 
  $U_{\mathrm{TTF}}\approx 6.2$ eV, 
while the Coulomb 
and exchange interactions between the L and R fragments are
  $V_0=2.07$ eV and $J=0.18$ eV.
Let us stress that $J_\mathrm{H}$ is not directly reduced to the
exchange coupling between the L and R moieties.
For the inter-molecular interactions,
there are three kinds of transfer integrals,
  $t_1$, $t_2$, and $t_3$ [shown in Fig.\ \ref{fig:fig2}(d)], and
 corresponding Coulomb repulsions, $V_1$, $V_2$,
  and $V_3$.
The correspondence relations
between the transfer integrals in the  MO basis 
  and those in the fragment basis are given by
$ t_1 =
(- t_{\mathrm{gg}} + t_{\mathrm{uu}} + 2 t_{\mathrm{gu}})/2$,
$ t_2 =
 (t_{\mathrm{gg}} + t_{\mathrm{uu}} )/2$,
and 
$t_3 =
 (- t_{\mathrm{gg}} + t_{\mathrm{uu}} - 2 t_{\mathrm{gu}})/2$.
The dominant transfer integrals ($>0.01$ eV) are 
$t_0=-0.17$,
  $t_1^{[001]}=-0.26$, $t_2^{[001]}=-0.17$,
  $t_1^{[012]} = -0.05$, and $t_1^{[013]}=0.02$, where 
  all energies are given in eV.
By fitting the GGA band calculation, 
these 
are estimated as
  $t_0 = -0.12$,
  $t_1^{[001]} = -0.20$, $t_2^{[001]} = -0.14$, 
  $t_1^{[012]} = -0.03$, and $t_1^{[013]}=0.01$.
The fitted band structure is shown by the thin solid lines 
  in Fig.\ \ref{fig:fig3}.
We obtain qualitatively consistent 
 parameters, and it is  worth noting 
that 
 the inter-fragment transfer integral within
 the molecule $|t_0|$  
  is smaller than the inter-molecular transfer integral
 $|t_1^{[001]}|$.
A similar situation has been pointed out in 
single-component molecular metals. 
\cite{Ishibashi-Seo-SCMM}

\begin{figure}[t]
\begin{center}
\includegraphics[width=6.5cm]{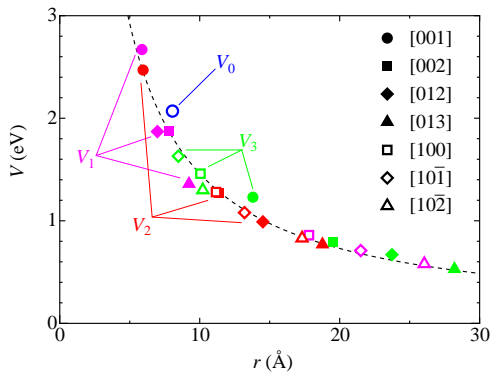}
\end{center}
\caption{
(Color online)
Distance dependence of the inter-fragment Coulomb repulsions.
The dotted line represents the bare Coulomb repulsion.
The Coulomb repulsion within the fragment is $U=5.30$ eV.
}
\label{fig:fig4}
\end{figure}

We have also calculated 
the Coulomb repulsions as a function 
  of distance between the fragments.
We define the inverse of the inter-fragment distance, $1/r$,
as the average inverse distance
between 6 sulfur atoms in each fragment.
The magnitudes of interactions follow 
 surprisingly well the $1/r$ Coulomb
  law, including the intra-molecular interaction $V_0$, 
 as shown in Fig.\ \ref{fig:fig4}.
This result strongly supports our fragment
 decomposition picture. 
In conventional single-orbital compounds,
  the bare Coulomb 
  interactions are known to follow the Coulomb law
 as a function of the inter-molecular distance.
\cite{Mori2000,Nakamura2009} 
In addition, we find that
  the inter-fragment distance for the [001] interactions
   becomes shorter
  than that for the intra-molecular one, and then
the interactions $V_1^{[001]}$ and $V_2^{[001]}$  exceed the
  intra-molecular interaction  $V_0$.
This feature
would be a key ingredient in obtaining the intra-molecular
  degree of freedom, and particularly for 
  the ICO phenomena. 
However, we should note that
   the screening effects
  in the crystal are not taken into account.
Such  effects have been examined in the single-orbital system,
 \cite{Cano-Cort2007,Nakamura2009}
and it remains future work to extend the analysis to 
  the multi-orbital systems.

We briefly discuss the electronic states in (TTM-TTP)I$_3$
on the basis of the present fragment-MO picture.
From the simple "atomic" limit analysis of our model,
 where the kinetic energy is neglected,
we indeed find that 
  the lowest-energy state is the 
  ICO state in which 
  the charge is disproportionated at the R (L) fragment 
  in molecule $\I$ ($\II$) 
  in Fig.\ \ref{fig:fig2}(d).
This ICO pattern is compatible with 
  the $\bm q =(0,0,1/2)$  superstructure observed by
 the X-ray measurements.
\cite{Maesato1999,Fujimura1999,Nogami2003} 
On the basis of this finding,
we infer that
the non-magnetic insulating behavior observed at low temperatures
  \cite{Maesato1999,Fujimura1999,Onuki2001} 
 is attributed to the spin-singlet formation
  on the $t_1^{[001]}$ bond with two-fold periodicity along the stacking
  direction.
On the other hand,
 the paramagnetic non-metallic state 
seen at high temperature 
\cite{Mori1994,Mori1997,Yasuzuka2006}
  might be due to the charge localization on each 
  $t_1^{[001]}$ bond, since  $t_1^{[001]}$ is the strongest bond in the system.
A detailed analysis of possible ordered states 
based on the derived effective model 
will be published elsewhere. 

In summary,
we have constructed 
an effective  Hubbard-type Hamiltonian
  and examined the electronic band structure for 
  the multi-orbital molecular compound (TTM-TTP)I$_3$.
It has been clarified that the present scheme combined with 
  the \textit{ab initio} calculations 
is consistent with
  the GGA calculation.
We have found that, in the fragment picture, 
  the stacked TTM-TTP molecules can be described by the effective 
 two-leg ladder and the Coulomb repulsions are 
proportional to the inverse of the inter-fragment
distance.

\acknowledgements
MT thanks S.\ Yasuzuka, T.\ Kawamoto, T.\ Mori, and K.\ Yakushi 
  for fruitful discussions on the experimental aspects of 
  (TTM-TTP)I$_3$.
MT and YO also thank L.\ Cano-Cort\'es, J. Merino, and K.\ Nakamura
   for discussions
  on the parameter evaluations of the molecular solids.
YO and MLB were supported by the Grant-in-Aid for JSPS Fellows.
This research was partially supported by
Grants-in-Aid for Scientific Research on Innovative Areas 
(20110002, 20110003, and 20110004)
from the 
Ministry of Education, Culture, Sports, Science and Technology, Japan.

\end{document}